\documentclass[aps,pre,showpacs,groupedaddress]{revtex4}
\usepackage{amsfonts}
\usepackage{mathrsfs}
\usepackage{dcolumn}
\usepackage{graphicx}
\usepackage{amsmath}
\usepackage{amssymb}
\usepackage{amsbsy}

\begin{document}

\title{Origin of the computational hardness for learning with binary synapses}

\author{Haiping Huang}
\affiliation{Department of Computational
Intelligence and Systems Science, Tokyo Institute of Technology,
Yokohama 226-8502, Japan}
\author{Yoshiyuki Kabashima}
\affiliation{Department of
Computational Intelligence and Systems Science, Tokyo Institute of
Technology, Yokohama 226-8502, Japan}
\date{\today}

\begin{abstract}
Supervised learning in a binary perceptron is able to classify an
extensive number of random patterns by a proper assignment of binary
synaptic weights. However, to find such assignments in practice, is
quite a nontrivial task. The relation between the weight space
structure and the algorithmic hardness has not yet been fully
understood. To this end, we analytically derive the Franz-Parisi
potential for the binary preceptron problem, by starting from an
equilibrium solution of weights and exploring the weight space
structure around it. Our result reveals the geometrical organization
of the weight space\textemdash the weight space is composed of
isolated solutions, rather than clusters of exponentially many
close-by solutions. The point-like clusters far apart from each
other in the weight space explain the previously observed glassy
behavior of stochastic local search heuristics.

\end{abstract}

\pacs{89.75.Fb, 87.19.L-, 75.10.Nr}
 \maketitle

\section{Introduction}
To provide an analytic explanation for
general phenomena using simple theoretical concept is the most
interesting part of physics. Statistical physics methods in spin
glass theory provide new tools and ideas to study many hard
constraint satisfaction problems~\cite{MM-2009}, especially the
relation between detailed organization of solutions in the solution
space and the algorithmic hardness~\cite{Krzakala-PNAS-2007}.

A prototypical example is the binary perceptron problem, where $N$
input neurons (units) are connected to a single output unit by
synapses of binary value ($\pm1$) synaptic weights. These weights
have to be inferred from a set of examples (input patterns) with
desired classification labels (supervised learning). An assignment
of these weights is referred to as a solution if the perceptron
manages to classify all the input patterns by this assignment. The
ratio between the number of patterns and the number of synapses is
called the constraint density. Each example acts as a constraint on
the solution space, since increasing the number of examples causes
the shrinkage of the space. The critical constraint density was
reported to be about $0.833$~\cite{Krauth-1989}, below which the
solution space is typically nonempty.

The binary perceptron serves as an elementary building block of
complex neural networks and is also one of the basic structures for
learning and memory~\cite{Engel-2001}. Memory in neuronal systems is
stored in the synaptic weights, and a binary synaptic weight is
robust against noise and also suitable for simple hardware
implementation in applications. The binary perceptron has thus a
wide variety of applications ranging from rule inference or
structure mining in machine learning~\cite{Engel-2001} to error
correcting codes or data compression in information
theory~\cite{Hosaka-2002}, and even high-dimensional data analysis
in biology~\cite{Weigt-09}. However, a learning task in the binary
perceptron is known to be an NP(nondeterministic polynomial
time)-complete problem in the worst case~\cite{Blum-1992}. Many
efforts have been devoted to design low-complexity algorithms to
find a solution for a typical case of this difficult
problem~\cite{Kohler-1990,Patel-1993,Bouten-1998,Zecchina-2006,Kaba-08,Huang-2010jstat,Huang-2011epl,Saad-2013}.
However, for many local search heuristics, the search process slows
down as the constraint density grows, and the learning threshold
decreases as the number of synapses
increases~\cite{Patel-1993,Huang-2010jstat,Huang-2011epl}. This
typical glassy behavior of stochastic local search algorithms
remains to be explained and was conjectured to be related to the
geometrical organization of the solution
space~\cite{Horner-1992a,Huang-2011epl,Kaba-09,Huang-JPA2013}. The
statistical properties of this problem were intensively studied by the
statistical physics community in the past
decades~\cite{Gardner-1988b,Krauth-1989,Kaba-09,Engel-2001}.
However, an analytic computation of a conclusive picture of the
solution space structure is still lacking so far, although this is
an important topic both in computer science (machine learning or
computational neuroscience) and in statistical physics.

A recent study~\cite{Huang-JPA2013} carried out an entropy landscape
analysis by focusing on the solution-pairs separated by certain
Hamming distance (the number of elements in different states in two
solutions), which motivated us to propose a suitable and solid
framework to provide a comprehensive description of the solution
space. The basic idea is to select an equilibrium solution sampled
from the Boltzmann measure, and then explore the solution space
around this selected equilibrium solution by analyzing the entropy
landscape in the vicinity of the reference equilibrium solution.
This framework was originally introduced as the name of Franz-Parisi
potential to study the metastable state structure for discontinuous
mean-field spin glasses (e.g., $p$-spin spherical spin
glass)~\cite{Franz-1995,Franz-1997prl,Franz-1998}, where the
potential has the physical meaning of the free energy cost to keep a
system at a temperature with a fixed overlap from an equilibrium
configuration at a different temperature. In this work, the
Franz-Parisi potential is interpreted in terms of the entropy
function to describe the solution space, and we show that a quenched
computation (average over the choice of the reference equilibrium
solution) of the potential in the zero temperature limit is possible
and provides important physical insights towards understanding the
geometrical organization of the solution (weight) space.

Our computation demonstrates that the weight space of the binary
perceptron problem is indeed made of isolated solutions for any
finite constraint density, with the minimal Hamming distance
separating two solutions growing with the constraint density. This
study reveals the origin of the computational hardness in the binary
perceptron problem, explaining the known fact that when the number
of synapses becomes sufficiently large, an exponential scaling in
computational time is required to maintain a fixed finite constraint
density for a learning
task~\cite{Horner-1992a,Huang-2011epl,Patel-1993}.

In Sec.~\ref{model}, we define in detail the binary perceptron problem. In Sec.~\ref{compt},
we introduce the Franz-Parisi potential framework and derive the explicit form
of the potential under the replica symmetric approximation. Results are presented and discussed in
Sec.~\ref{Disc}. Concluding remarks and future perspectives are given in Sec.~\ref{Conc}.
\section{The binary perceptron problem}
\label{model}
 The binary perceptron is a
single-layered feed-forward neural network, i.e., $N$ input neurons
are connected to a single output neuron by $N$ synapses of weight
$J_i = \pm 1$ $(i = 1, 2, \ldots, N)$. The perceptron tries to learn
$P = \alpha N$ associations $\{\boldsymbol{\xi}^\mu , \sigma_0^\mu
\}$ $(\mu= 1, 2, \ldots, P)$, where $\boldsymbol{\xi}^{\mu} \equiv
(\xi_1^\mu, \xi_2^\mu, \ldots, \xi_N^\mu)$ is an input pattern with
$\xi_i^\mu = \pm 1$, and $\sigma_0^\mu = \pm 1$ is the desired
classification of the input pattern $\mu$. For a random
classification task,  both $\{\xi_{i}^{\mu}\}$ and the desired
output $\{\sigma_{0}^{\mu}\}$ are generated randomly independently
with $\xi_i^{\mu}$ and $\sigma_0^\mu$ being $\pm 1$ with probability
$1/2$. Given the input pattern $\boldsymbol{\xi}^{\mu}$, the actual
output $\sigma^{\mu}$ of the perceptron is $\sigma^{\mu}={\rm
sgn}\left(\sum_{i=1}^{N}J_{i}\xi_{i}^{\mu}\right)$. If
$\sigma^{\mu}=\sigma_{0}^{\mu}$, we say that the synaptic weight
vector $\mathbf{J}$ has learned the $\mu$-th pattern. Each input
pattern imposes a constraint on all synaptic weights, therefore
$\alpha$ denotes the constraint density. The solution space of the
binary perceptron is composed of all the weight configurations
$\{J_i\}$ that satisfy  $\sigma_0^\mu \sum_i J_i \xi_i^\mu > 0$ for
$\mu = 1, 2, \ldots, P$. The energy cost is thus defined as the
number of patterns mapped
incorrectly~\cite{Engel-2001,Huang-JPA2013}, i.e.,
\begin{equation}\label{energy}
E(\mathbf{J})=\sum_{\mu}\Theta\left(-\frac{\sigma_{0}^{\mu}}{\sqrt{N}}\sum_{i=1}^{N}J_{i}\xi_{i}^{\mu}\right),
\end{equation}
where $\Theta(x)$ is a step function with the convention that
$\Theta(x)=0$ if $x\leq0$ and $\Theta(x)=1$ otherwise. The prefactor
$N^{-1/2}$ is introduced to ensure that the argument of the step
function remains at the order of unity, for the sake of the
following statistical mechanical analysis in the thermodynamic
limit. Without loss of generality, we assume $\sigma_{0}^{\mu}=+1$
for any input pattern in the remaining part of this paper, since one
can perform a gauge transformation
$\xi_{i}^{\mu}\rightarrow\xi_{i}^{\mu}\sigma_{0}^{\mu}$ to each
input pattern without affecting the result.

From a theoretical perspective, the perceptron is typically able to
learn an extensive number of random input patterns with the storage
capacity $\alpha_s\simeq0.833$~\cite{Krauth-1989}. However, to find
such a solution configuration $\mathbf{J}$ in practice, is quite a
nontrivial task. Here, to reveal the origin of this computational
hardness, we apply the replica method from the theory of disordered
systems~\cite{MM-2009} to derive an analytic expression of the
Franz-Parisi potential, which characterizes the entropy landscape of
the problem.

\section{Analytic computation of the Franz-Parisi potential}
\label{compt}
The
 binary perceptron problem is a densely-connected graphical model~\cite{Huang-JPA2013} in that a proper
 assignment of all synaptic weights is needed to satisfy each constraint (learn each pattern). Its equilibrium
 property can thus be described by mean-field computation in terms
 of the Franz-Parisi potential.
The basic idea is to first select an equilibrium configuration
$\mathbf{J}$ at a temperature $T'$, then constrain its overlap with
another equilibrium configuration $\mathbf{w}$ at a different
temperature $T$, which yields a constrained
free energy~\cite{Franz-1995}:
\begin{equation}\label{free-energy}
F(T,T',x)=\left<\frac{1}{Z(T')}\sum_{\mathbf{J}}e^{-\beta'E(\mathbf{J})}\ln\sum_{\mathbf{w}}e^{-\beta
E(\mathbf{w})+x\mathbf{J}\cdot\mathbf{w}}\right>,
\end{equation}
after taking the quenched disorder average (over the pattern
distribution $\boldsymbol{\xi}$, denoted by the angular bracket) and
the average over the distribution of $\mathbf{J}$, which is
$e^{-\beta'E(\mathbf{J})}/Z(T')$. $Z(T')$ is the partition function
for the original measure and $\beta(\beta')$ is the inverse
temperature. The constrained free energy
$\ln\sum_{\mathbf{w}}e^{-\beta
E(\mathbf{w})+x\mathbf{J}\cdot\mathbf{w}}$ is a self-averaging
quantity with respect to both the quenched disorder and the
probability distribution of the reference configuration
$\mathbf{J}$~\cite{Franz-1997prl}. Its value doesnot depend on the
particular realization and coincides with the typical value, which
can be calculated via the replica method.

In our current setting, we are interested in the ground states of
the problem, thus we set $\beta=\beta'\rightarrow\infty$, arriving
at the following formula:
\begin{equation}\label{replica}
 F(x)=\lim_{\substack{n\rightarrow 0\\m\rightarrow
0}}\frac{\partial}{\partial
m}\left<\sum_{\{\mathbf{J}^{a},\mathbf{w}^{\gamma}\}}\prod_{\mu}\left[\prod_{a,\gamma}\Theta(u_{a}^{\mu})\Theta(v_{\gamma}^{\mu})\right]
e^{x\sum_{\gamma,i}J_{i}^{1}w_{i}^{\gamma}}\right>,
\end{equation}
where $u_{a}^{\mu}\equiv\sum_{i}J_{i}^{a}\xi_{i}^{\mu}/\sqrt{N}$ and
$v_{\gamma}^{\mu}\equiv\sum_{i}w_{i}^{\gamma}\xi_{i}^{\mu}/\sqrt{N}$.
In Eq.~(\ref{replica}), we have $n$ replicas
$\mathbf{J}^{a}(a=1,\ldots,n)$ and $m$ replicas
$\mathbf{w}^{\gamma}(\gamma=1,\ldots,m)$, with the coupling field
($x$) term being an interaction of all the constrained replicas
$\mathbf{w}^{\gamma}$ with one privileged replica $\mathbf{J}^{1}$.
The replica method to compute the typical value of the constrained
free energy is based on two mathematical identities: $\ln
Z=\lim_{m\rightarrow 0}\frac{\partial Z^{m}}{\partial m}$ and
$Z^{-1}=\lim_{n\rightarrow0}Z^{n-1}$. To evaluate the average in
Eq.~(\ref{replica}), we need to define the overlap matrixes
$Q_{ab}\equiv\mathbf{J}^{a}\cdot\mathbf{J}^{b}/N$,
$P_{a\gamma}\equiv\mathbf{J}^{a}\cdot\mathbf{w}^{\gamma}/N$ and
$R_{\gamma\eta}\equiv\mathbf{w}^{\gamma}\cdot\mathbf{w}^{\eta}/N$,
which characterize the following disorder averages
$\left<u_{a}^{\mu}u_{b}^{\mu}\right>=Q_{ab}$,
$\left<u_{a}^{\mu}v_{\gamma}^{\mu}\right>=P_{a\gamma}$ and
$\left<v_{\gamma}^{\mu}v_{\eta}^{\mu}\right>=R_{\gamma\eta}$. Under
the replica symmetric (RS) ansatz, we have
$Q_{ab}=q(1-\delta_{ab})+\delta_{ab}$,
$P_{a\gamma}=p\delta_{a1}+p'(1-\delta_{a1})$ and
$R_{\gamma\eta}=r(1-\delta_{\gamma\eta})+\delta_{\gamma\eta}$, where
$\delta_{ab}=1$ if $a=b$ and $0$ otherwise.

 After some algebraic
manipulations, we finally get the constrained free energy density
$f(x)$ as:
\begin{widetext}
\begin{equation}\label{VFP}
\begin{split}
    f(x) &=\lim_{N\rightarrow\infty}F(x)/N=\frac{\hat{r}}{2}(r-1)-p\hat{p}+p'\widehat{p'}+xp+\alpha\int D\omega\int DtH^{-1}(\tilde{t})\int_{\tilde{t}}^{\infty}Dy\ln H(h(\omega,t,y))\\
    &+\int
    D\mathbf{z}(2\cosh\hat{a})^{-1}\Biggl[e^{\hat{a}}\ln2\cosh(\hat{a}'+\hat{p}-\widehat{p'})+e^{-\hat{a}}\ln2\cosh(\hat{a}'-\hat{p}+\widehat{p'})\Biggr],
    \end{split}
\end{equation}
\end{widetext}
where $\int D\mathbf{z}\equiv\int Dz_{1}Dz_{2}Dz_{3}$,
$\tilde{t}\equiv-\sqrt{\frac{q}{1-q}}t$, and
$H(x)\equiv\int_{x}^{\infty} Dz$ with the Gaussian measure $Dz\equiv
G(z)dz$ in which $G(z)\equiv\exp(-z^{2}/2)/\sqrt{2\pi}$.
$h(\omega,t,y)\equiv-\left((p-p')y/\sqrt{1-q}+\sqrt{v_{\omega}}\omega+p't/\sqrt{q}\right)/\sqrt{1-r}$
where $v_{\omega}\equiv r-p'^{2}/q-(p-p')^{2}/(1-q)$.
$\hat{a}\equiv\sqrt{\hat{q}-\widehat{p'}}z_{1}+\sqrt{\widehat{p'}}z_{3}$
and
$\hat{a}'\equiv\sqrt{\hat{r}-\widehat{p'}}z_{2}+\sqrt{\widehat{p'}}z_{3}$.
The associated self-consistent (saddle-point) equations for the
order parameters $\{q,\hat{q},r,\hat{r},p,\hat{p},p',\widehat{p'}\}$
are derived in the Appendix~\ref{app:CFE}.

The Franz-Parisi potential $\mathcal {V}(p)$ is obtained through a
Legendre transform of $f(x)$, i.e., $\mathcal {V}(p)=f(x)-xp$ and
$\frac{{\rm d}f(x)}{{\rm d}x}=p$. $\mathcal {V}(p)$ has the meaning
of the entropy characterizing the growth rate of the number of
solutions ($e^{N\mathcal {V}(p)}$) lying apart at a normalized
distance $(1-p)/2$ (Hamming distance divided by $N$) from the fixed
equilibrium solution. Detailed information about the solution space
structure can be extracted from the behavior of this potential at
different values of $p$, especially those values close to one. Since
the potential curve may lose its concavity, one has to solve
numerically the saddle-point equations (see Appendix~\ref{app:Deriv}) by fixing $p$ and
searching for compatible coupling field $x$ (by using the secant
method~\cite{NumOpt}).


\begin{figure}
          \includegraphics[bb=16 20 289 219,scale=0.85]{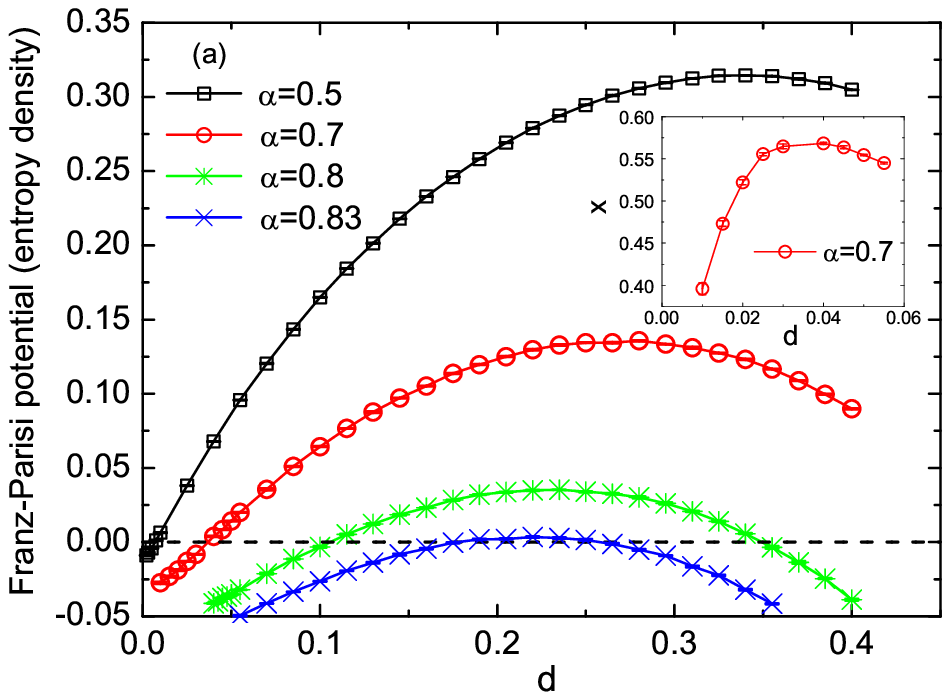}
     \hskip .1cm
     \includegraphics[bb=14 23 289 222,scale=0.85]{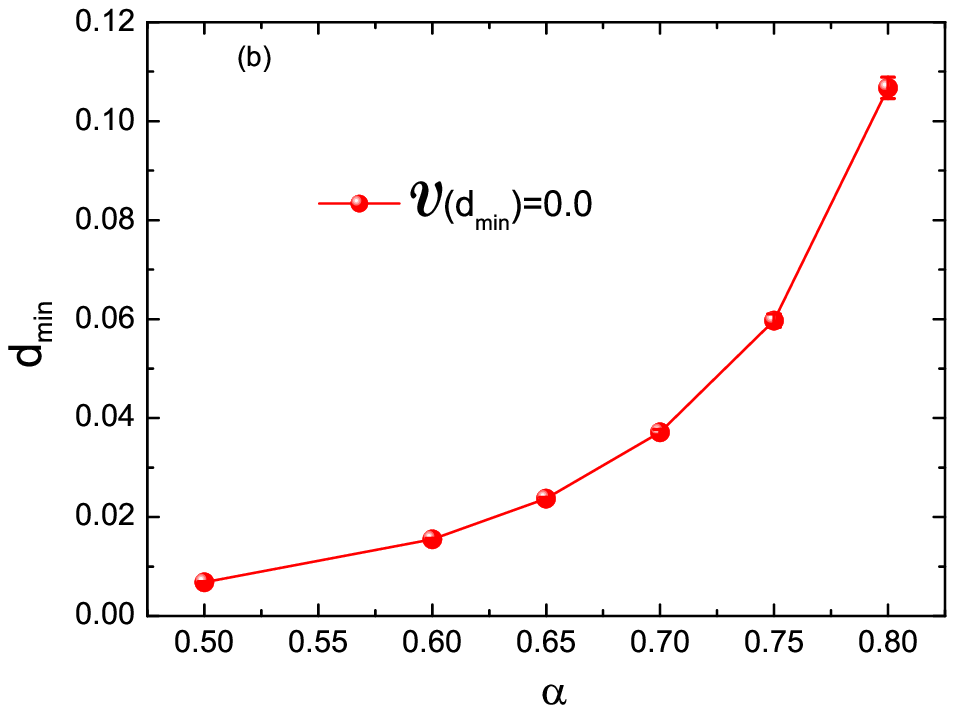}
 \vskip .1cm
     \includegraphics[bb=95 616 351 759,scale=0.9]{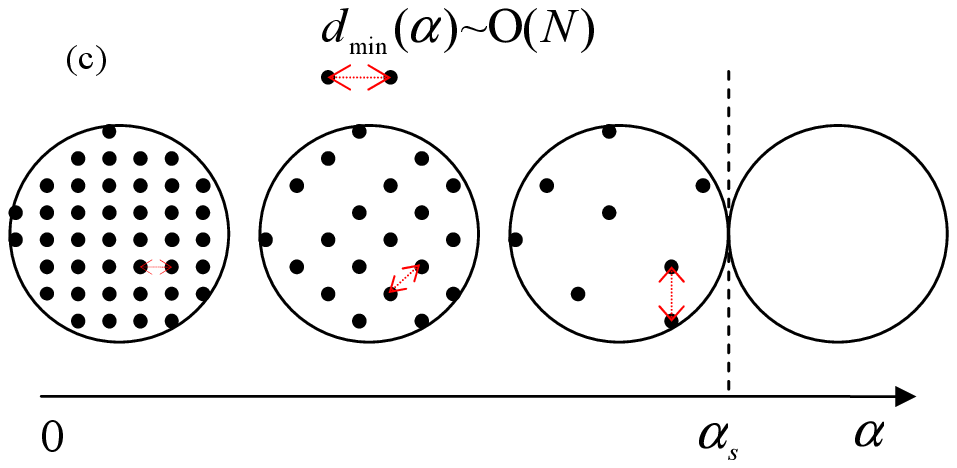}
     \vskip .1cm
  \caption{(Color online)
     Entropy landscape of solutions in the binary perceptron problem. Iterations of the saddle-point
      equations are always converged to produce the data points. The error bars give statistical errors and are smaller than or
      equal to the symbol size. (a) Franz-Parisi potential as a function of the normalized Hamming distance. The behavior of the
     coupling field with the distance is shown in the inset for $\alpha=0.7$, for which an observed maximum implies the change of
     the concavity of the entropy curve (this also holds for other finite values of $\alpha$).
      (b) Minimal distance versus the constraint density. Within the
      minimal distance, there are no solutions satisfying the
      distance constraint from the reference equilibrium solution.
      (c) Schematic illustration of the weight space based on results of (a)
      and (b). The points indicate the equilibrium solutions of
      weights. $\alpha_s\simeq0.833$ is the storage capacity after which the solution space is typically
      empty. $d_{\rm min}$ is the actual Hamming
distance without normalization.}\label{V-FP}
 \end{figure}

 \section{Solution space consists of isolated solutions}
 \label{Disc}
The
Franz-Parisi potential versus the predefined normalized Hamming
distance ($d=(1-p)/2$) is shown in Fig.~\ref{V-FP} (a). At the
maximum corresponding to $x=0$ ($x=-\frac{{\rm d}\mathcal {V}}{{\rm
d}p}=\frac{1}{2}\frac{{\rm d}\mathcal {V}}{{\rm d}d}$),
$\mathcal {V}(p)$ gives back the entropy of the original system. As
the distance gets close to zero, one finds that there exists a value
of distance at which the entropy curve loses its concavity and turns
to a convex part (see the inset of Fig.~\ref{V-FP} (a) and note that
the sign of the slope changes at the maximum point). This behavior
leads to an important result that there exists a minimal distance of
$\mathcal {O}(N)$ below which no solutions are separated from the
reference equilibrium solution. Note that the reference solution is
distributed according to the Boltzmann measure (a uniform measure
over all solutions). The minimal distance grows with the constraint
density, as shown in Fig.~\ref{V-FP} (b). This can be understood by
the following argument. Due to the hard nature of the pattern
constraint in the binary perceptron problem---all synapses are
involved in classifying each input pattern, flipping one synaptic
weight should force the rearrangement of many weight values to
memorize the learned patterns. Similar phenomena were also observed
in Gallager's type error correcting code~\cite{Montanari-2004} and
locked constraint satisfaction problem~\cite{Lenka-08prl}.

For small $\alpha$, it is not easy to show the convex part
numerically. However, one can prove that when $p\rightarrow 1$, the
Franz-Parisi potential vanishes as expected for all
$\alpha$ (see Appendix~\ref{app:Proof}). In addition, at $p\rightarrow 1$
($\epsilon\equiv1-p\rightarrow 0$), we have $\frac{{\rm d}\mathcal
{V}(p)}{{\rm d}p}=\alpha
C_{p}\epsilon^{-1/2}+(\ln\epsilon)/2+C$ (see Appendix~\ref{app:Deriv}) where $C$ is a
finite constant and $C_{p}$ is a positive constant. The first term
dominates the divergent behavior in the limit $\epsilon\rightarrow
0$. This means that, for any finite $\alpha>0$, the entropy curve in
Fig.~\ref{V-FP} (a) has a negative infinite slope ($\frac{{\rm
d}\mathcal {V}}{{\rm d}d}=-2\frac{{\rm d}\mathcal {V}}{{\rm d}p}$)
at $p=1$, supporting the existence of the convex part and the
minimal distance. As expected from the tendency shown in
Fig.~\ref{V-FP} (b), the value of the minimal distance becomes very
small for the less constrained case (small constraint density). This
explains why a simple local search algorithm can find a solution
when either $N$ or $\alpha$ is
small~\cite{Kohler-1990,Patel-1993,Bouten-1998,Huang-2010jstat,Huang-2011epl,Saad-2013}.
As $\alpha$ increases, the minimal distance grows rapidly, as a
consequence, any algorithms working by local move (each time a few
weights are flipped) should find increasing difficulty to identify a
solution (especially at a very large $N$), which holds even for
reinforced message passing algorithms~\cite{Zecchina-2006}. In other
words, an extensive energy or entropic barrier should be overcome.
The energy landscape is always valleys dominated (valleys are
metastable states with positive energy cost). These metastable
states are much more numerous than the frozen ground
states~\cite{Lenka-2010prb}. Local algorithms will get trapped by
these metastable states with high probability.

We thus conclude that, at variance with random $K$-SAT or
$Q$-coloring problems~\cite{Krzakala-PNAS-2007}, the solution space
of the binary perceptron problem is simple in the sense that it is
made of isolated solutions instead of well separated clusters of
exponentially many close-by solutions. This picture is consistent
with evidences reported in previous
studies~\cite{Uezu-1995,Kaba-09,Huang-JPA2013}. Moreover,
non-convergence of the iteration of the saddle-point equations was
not observed, which may be related to the simple structure of the
solution space. In fact, below the storage capacity, the replica
symmetric solution is stable without any need to introduce replica
symmetry breaking scenario for this
problem~\cite{Gardner-1988b,Krauth-1989}. Our quenched computation
of the Franz-Parisi potential reveals that, synaptic weights to
realize the random classification task are organized into point-like
clusters (zero internal entropy) far apart from each other (see
Fig.~\ref{V-FP} (c)), with the result that in the thermodynamic
limit, an exponential computation time is required to reach a finite
fixed $\alpha$~\cite{Horner-1992a,Patel-1993}.

\section{Conclusion}
\label{Conc}
We give an analytic expression of the
Franz-Parisi potential for the binary perceptron problem. This
potential describes the entropy landscape of solutions in the
vicinity of a reference equilibrium solution, and its shape is
independent of the choice of the reference point. Solving the
saddle-point equations, we find that the concavity of the curve
changes at some distance, leading to a minimal distance below which
there doesnot exist solutions satisfying the distance constraint.
Furthermore, this minimal distance increases with the constraint
density, implying that the problem is extremely hard because the
solution space is composed of isolated solutions (point-like
clusters) with the property that to go from one solution to another
solution, one should flip an extensive number (proportional to $N$)
of synaptic weights.

Our analysis establishes a refined picture of the organization
structure of the solution space for the binary perceptron problem,
which is helpful for understanding the glassy behavior of local
search heuristics~\cite{Patel-1993,Huang-2010jstat,Huang-2011epl},
which may have some connections with recent studies of constrained
glasses~\cite{Franz-2013}, and furthermore, is expected to shed
light on design of efficient algorithms for large-scale neuromorphic
devices. The analytic analysis presented in this paper also offers a
basis for possible rigorous mathematical (probabilistic) analysis of
the entropy landscape~\cite{Nature-2005}, and has potentially applications for studying the solution space structure of other hard
problems in information processing, e.g., spike time-based neural
classifiers~\cite{tempotron-2006nature,tempotron-2010prl,Baldassi-2013}.


\begin{acknowledgments}
We thank Lenka Zdeborov\'a for helpful discussions and Haijun Zhou for helpful comments on the
manuscript. This work was
partially supported by the JSPS Fellowship for Foreign Researchers
(Grant No. $24\cdot02049$) (H.H.) and JSPS/MEXT KAKENHI Grant No.
$25120013$ (Y.K.). Support from the JSPS Core-to-Core Program
\textquotedblleft Non-equilibrium dynamics of soft matter and
information\textquotedblright is also acknowledged.
\end{acknowledgments}
\appendix
\section{Derivation of constrained free energy}
\label{app:CFE}
In the current context, for a reference equilibrium configuration
$\mathbf{J}$ at temperature $T'$, one is interested in the free
energy of a perturbed system (with the constraint that the
configuration $\mathbf{w}$  at temperature $T$ should satisfy a
prefixed overlap with $\mathbf{J}$), leading to the constrained free
energy~\cite{Franz-1995}:
\begin{equation}\label{free-energy}
F(T,T',x)=\left<\frac{1}{Z(T')}\sum_{\mathbf{J}}e^{-\beta'E(\mathbf{J})}\ln\sum_{\mathbf{w}}e^{-\beta
E(\mathbf{w})+x\mathbf{J}\cdot\mathbf{w}}\right>_{\boldsymbol{\xi}},
\end{equation}
where $Z(T')=\sum_{\mathbf{J}}e^{-\beta'E(\mathbf{J})}$ and $x$ is
the coupling field to control the overlap (or distance) between two
configurations, i.e., $p\equiv\mathbf{J}\cdot\mathbf{w}/N$. We are
interested in the ground state, then we set both inverse
temperatures equal and make them tend to infinity. Substituting the
definition of energy cost of the problem, and using
$e^{-\beta\Theta(-u)}=\Theta(u)$ in the zero temperature limit, we
have
\begin{equation}\label{free-energy02}
F(x)=\left<\frac{1}{Z(T')}\sum_{\mathbf{J}}\Theta\left(\frac{1}{\sqrt{N}}\sum_{i=1}^{N}J_{i}\xi_{i}^{\mu}\right)
\ln\sum_{\mathbf{w}}\Theta\left(\frac{1}{\sqrt{N}}\sum_{i=1}^{N}w_{i}\xi_{i}^{\mu}\right)e^{x\mathbf{J}\cdot\mathbf{w}}\right>_{\boldsymbol{\xi}}.
\end{equation}

To evaluate the typical value of $F(x)$, we resort to the replica
method~\cite{Engel-2001}, by using two mathematical identities: $\ln
Z=\lim_{m\rightarrow 0}\frac{\partial Z^{m}}{\partial m}$ and
$Z^{-1}=\lim_{n\rightarrow0}Z^{n-1}$. Introducing $n$ unconstrained
replicas $\mathbf{J}^{a}(a=1,\ldots,n)$ and $m$ constrained replicas
$\mathbf{w}^{\gamma}(\gamma=1,\ldots,m)$, we rewrite $F(x)$ as:
\begin{equation}\label{replica}
 F(x)=\lim_{\substack{n\rightarrow 0\\m\rightarrow
0}}\frac{\partial}{\partial
m}\left<\sum_{\{\mathbf{J}^{a},\mathbf{w}^{\gamma}\}}\prod_{\mu}\left[\prod_{a,\gamma}\Theta(u_{a}^{\mu})\Theta(v_{\gamma}^{\mu})\right]
e^{x\sum_{\gamma,i}J_{i}^{1}w_{i}^{\gamma}}\right>_{\boldsymbol{\xi}},
\end{equation}
where $u_{a}^{\mu}\equiv\sum_{i}J_{i}^{a}\xi_{i}^{\mu}/\sqrt{N}$ and
$v_{\gamma}^{\mu}\equiv\sum_{i}w_{i}^{\gamma}\xi_{i}^{\mu}/\sqrt{N}$.
To proceed, we define the following overlap matrixes:
$Q_{ab}\equiv\mathbf{J}^{a}\cdot\mathbf{J}^{b}/N$,
$P_{a\gamma}\equiv\mathbf{J}^{a}\cdot\mathbf{w}^{\gamma}/N$ and
$R_{\gamma\eta}\equiv\mathbf{w}^{\gamma}\cdot\mathbf{w}^{\eta}/N$,
which characterize the following disorder averages
$\left<u_{a}^{\mu}u_{b}^{\mu}\right>=Q_{ab}$,
$\left<u_{a}^{\mu}v_{\gamma}^{\mu}\right>=P_{a\gamma}$ and
$\left<v_{\gamma}^{\mu}v_{\eta}^{\mu}\right>=R_{\gamma\eta}$. By
inserting delta functions for these definitions and using their
integral representations, we obtain the disorder average $\mathcal
{S}$ in Eq.~(\ref{replica}) as:
\begin{equation}\label{replica02}
\begin{split}
    \mathcal {S}&=\prod_{a<b}\prod_{\gamma<\eta}\prod_{a,\gamma}\int\frac{dQ_{ab}d\hat{Q}_{ab}}{2\pi}\int\frac{dR_{\gamma\eta}d\hat{R}_{\gamma\eta}}{2\pi}\int\frac{dP_{a\gamma}d\hat{P}_{a\gamma}}{2\pi}
    e^{-{\rm
    i}\left(\sum_{a<b}Q_{ab}\hat{Q}_{ab}+\sum_{\gamma<\eta}R_{\gamma\eta}\hat{R}_{\gamma\eta}+\sum_{a,\gamma}P_{a\gamma}\hat{P}_{a\gamma}\right)}\\
    &\times\sum_{\{\mathbf{J}^{a},\mathbf{w}^{\gamma}\}}e^{\frac{{\rm i}}{N}\left(\sum_{a<b}\hat{Q}_{ab}\sum_{i}J_{i}^{a}J_{i}^{b}+\sum_{\gamma<\eta}\hat{R}_{\gamma\eta}\sum_{i}w_{i}^{\gamma}w_{i}^{\eta}
    +\sum_{a,\gamma}\hat{P}_{a\gamma}\sum_{i}J_{i}^{a}w_{i}^{\gamma}\right)}\\
    &\times\left<\prod_{\mu}\left[\prod_{a,\gamma}\Theta(u_{a}^{\mu})\Theta(v_{\gamma}^{\mu})\right]\right>_{\boldsymbol{\xi}}e^{x\sum_{i,\gamma}J_{i}^{1}w_{i}^{\gamma}}.
    \end{split}
\end{equation}

Now we re-scale the variable ${\rm
i}\hat{Q}_{ab}/N\rightarrow\hat{Q}_{ab}$ (this also applies for
other conjugated variables). We apply the replica symmetric
approximation~\cite{Engel-2001}, which assumes the permutation
symmetry of the overlap matrix. To be more precise,
$Q_{ab}=q(1-\delta_{ab})+\delta_{ab}$,
$P_{a\gamma}=p\delta_{a1}+p'(1-\delta_{a1})$ and
$R_{\gamma\eta}=r(1-\delta_{\gamma\eta})+\delta_{\gamma\eta}$, where
$\delta_{ab}=1$ if $a=b$ and $0$ otherwise. We first simplify
$\sum_{a,\gamma}\hat{P}_{a\gamma}J^{a}w^{\gamma}$ as:
\begin{equation}\label{replica04}
\begin{split}
   \sum_{a,\gamma}\hat{P}_{a\gamma}J^{a}w^{\gamma}&=\widehat{p'}\sum_{a,\gamma}J^{a}w^{\gamma}+(\hat{p}-\widehat{p'})\sum_{\gamma}J^{1}w^{\gamma}\\
   &=\frac{\widehat{p'}}{2}\left[\left(\sum_{a}J^{a}+\sum_{\gamma}w^{\gamma}\right)^{2}-\left(\sum_{a}J^{a}\right)^{2}-\left(\sum_{\gamma}w^{\gamma}\right)^{2}\right]
   +(\hat{p}-\widehat{p'})\sum_{\gamma}J^{1}w^{\gamma},
    \end{split}
\end{equation}
where the site index $i$ is dropped off since each $i$ shares the
same formula. Then we compute the disorder average as:
\begin{equation}\label{replica05}
  \left<\prod_{\mu}\left[\prod_{a,\gamma}\Theta(u_{a}^{\mu})\Theta(v_{\gamma}^{\mu})\right]\right>_{\boldsymbol{\xi}}=\left[\int D\omega\int Dt\int_{\tilde{t}}^{\infty}
    DyH^{m}(h(\omega,t,y))H^{n-1}(\tilde{t})\right]^{\alpha N},
\end{equation}
where $\tilde{t}\equiv-\sqrt{\frac{q}{1-q}}t$,
 and $H(x)\equiv\int_{x}^{\infty}
Dz$ with the Gaussian measure $Dz\equiv G(z)dz$ in which
$G(z)=\exp(-z^{2}/2)/\sqrt{2\pi}$.
$h(\omega,t,y)\equiv-\left((p-p')y/\sqrt{1-q}+\sqrt{v_{\omega}}\omega+p't/\sqrt{q}\right)/\sqrt{1-r}$
where $v_{\omega}\equiv r-p'^{2}/q-(p-p')^{2}/(1-q)$. In deriving
Eq.~(\ref{replica05}), we have parameterized
$u_{a}=\sqrt{1-q}y_{a}+\sqrt{q}t$ and
$v_{\gamma}=\sqrt{1-r}y'_{\gamma}+(p-p')y_{1}/\sqrt{1-q}+\sqrt{v_{\omega}}\omega+p't/\sqrt{q}$,
by using independent standard Gaussian random variables
$\{y_{a},t,y'_{\gamma},\omega\}$ of zero mean and unit variance. The
parameterization retains the covariance structure of
$\{u_a,v_\gamma\}$. The pattern index ($\mu$) is also dropped off
for the same reason. After a few algebraic manipulations, we obtain
\begin{equation}\label{replica03}
\begin{split}
    \mathcal
    {S}&=\exp\left[-\frac{N(n-1)n}{2}q\hat{q}-\frac{N(m-1)m}{2}r\hat{r}-mNp\hat{p}-N(n-1)mp'\widehat{p'}+Nxpm-\frac{Nn}{2}\hat{q}-\frac{Nm}{2}\hat{r}\right]\\
    &\times\exp\left[N\ln\int Dz_{1}\int Dz_{2}\int Dz_{3}\mathcal
    {A}(\hat{q},\hat{r},\hat{p},\widehat{p'},m,n)\right]\\
    &\times\exp\left[\alpha N\ln\int D\omega\int Dt\int_{\tilde{t}}^{\infty}
    DyH^{m}(h(\omega,t,y))H^{n-1}(\tilde{t})\right],
    \end{split}
\end{equation}
after approximating the integral in Eq.~(\ref{replica02}) by its
dominant part (a saddle point analysis in the large $N$ limit). To
derive Eq.~(\ref{replica03}), the Hubbard-Stratonovich
transformation was used. In Eq.~(\ref{replica03}), $\mathcal
{A}(\hat{q},\hat{r},\hat{p},\widehat{p'},m,n)\equiv(2\cosh\hat{a})^{n-1}\left[e^{\hat{a}}(2\cosh(\hat{a}'+\hat{p}-\widehat{p'}))^{m}+e^{-\hat{a}}(2\cosh(\hat{a}'-\hat{p}+\widehat{p'}))^{m}\right]$,
in which
$\hat{a}\equiv\sqrt{\hat{q}-\widehat{p'}}z_{1}+\sqrt{\widehat{p'}}z_{3}$
and
$\hat{a}'\equiv\sqrt{\hat{r}-\widehat{p'}}z_{2}+\sqrt{\widehat{p'}}z_3$.

The saddle point analysis (also called Laplace method) implies that
$\mathcal {S}$ should take its maximal value so that $\mathcal
{L}\equiv\ln\mathcal {S}$ should be extremized with respect to the
order parameters
$\{q,\hat{q},r,\hat{r},p,\hat{p},p',\widehat{p'}\}$. Keeping up to
the first order in $n$, the extremization with respect to $q$ and
$\hat{q}$ gives the self-consistent equations for $q$ and $\hat{q}$
(see Eqs.~(\ref{SDEa}) and~(\ref{SDEb})). As expected, their values
do not rely on other order parameters characterizing the property of
the constrained replicas. These two equations describe the
$\mathbf{J}$ system at equilibrium, and it should not be affected by
the $\mathbf{w}$ system which follows a perturbed distribution
depending on the reference solution $\mathbf{J}$. Finally, one can
readily get the constrained free energy density following the
definition given in Eq.~(\ref{replica}):
\begin{equation}\label{VFP}
\begin{split}
    f(x) &=\lim_{N\rightarrow\infty}F(x)/N=\frac{\hat{r}}{2}(r-1)-p\hat{p}+p'\widehat{p'}+xp+\alpha\int D\omega\int DtH^{-1}(\tilde{t})\int_{\tilde{t}}^{\infty}Dy\ln H(h(\omega,t,y))\\
    &+\int
    D\mathbf{z}(2\cosh\hat{a})^{-1}\Biggl[e^{\hat{a}}\ln2\cosh(\hat{a}'+\hat{p}-\widehat{p'})+e^{-\hat{a}}\ln2\cosh(\hat{a}'-\hat{p}+\widehat{p'})\Biggr],
    \end{split}
\end{equation}
together with the associated saddle-point equations:
\begin{subequations}\label{SDE}
\begin{align}
q&=\int Dz\tanh^{2}(\sqrt{\hat{q}}z),\label{SDEa}\\
\hat{q}&=\frac{\alpha}{1-q}\int
Dt\mathcal {R}^{2}(\tilde{t}),\label{SDEb}\\
p&=\int
    D\mathbf{z}(2\cosh\hat{a})^{-1}\left[e^{\hat{a}}\tanh(\hat{a}'+\hat{p}-\widehat{p'})-e^{-\hat{a}}\tanh(\hat{a}'-\hat{p}+\widehat{p'})\right],\label{SDEc}\\
    \hat{p}&=x+\frac{\alpha}{\sqrt{(1-q)(1-r)}}\int D\omega\int Dt \mathcal
    {R}(\tilde{t})\mathcal {R}(h(\omega,t,y=\tilde{t})),\\
    r&=\int
    D\mathbf{z}(2\cosh\hat{a})^{-1}\left[e^{\hat{a}}\tanh^{2}(\hat{a}'+\hat{p}-\widehat{p'})+e^{-\hat{a}}\tanh^{2}(\hat{a}'-\hat{p}+\widehat{p'})\right],\label{SDEe}\\
    \hat{r}&=\frac{\alpha}{1-r}\int D\omega\int DtH^{-1}(\tilde{t})\int_{\tilde{t}}^{\infty}Dy\mathcal
    {R}^{2}(h(\omega,t,y)),\label{SDEf}\\
    p'&=\int
    D\mathbf{z}(2\cosh\hat{a})^{-1}\left[e^{\hat{a}}\tanh\hat{a}\tanh(\hat{a}'+\hat{p}-\widehat{p'})+e^{-\hat{a}}\tanh\hat{a}\tanh(\hat{a}'-\hat{p}+\widehat{p'})\right],\\
    \widehat{p'}&=\frac{\alpha}{\sqrt{(1-q)(1-r)}}\int D\omega\int DtH^{-1}(\tilde{t})\mathcal
    {R}(\tilde{t})\int_{\tilde{t}}^{\infty}Dy\mathcal
    {R}(h(\omega,t,y)),
\end{align}
\end{subequations}
where $\int D\mathbf{z}\equiv\int Dz_{1}Dz_{2}Dz_{3}$, and $\mathcal
{R}(x)\equiv G(x)/H(x)$. In deriving these equations, we have used a
useful property of the Gaussian measure $\int Dz z\mathcal
{F}(z)=\int Dz\mathcal {F}'(z)$ where $\mathcal {F}'(z)$ is the
derivative of the function $\mathcal {F}(z)$ with respect to $z$.

To solve these saddle-point equations, for example,
Eq.~(\ref{SDEf}), one efficient way is to generate a random number
$y$ according to the conditional distribution ${\rm
Pr}(y|t)=\frac{G(y)\Theta(\sqrt{1-q}y+\sqrt{q}t)}{H(-\sqrt{\frac{q}{1-q}}t)}$
each time when using Monte-Carlo method to perform the integral. In
some cases, one may reexpress $\hat{a}$ and $\hat{a}'$ to retain
their covariances $\left<\hat{a}\hat{a}'\right>=\widehat{p'}$ (their
means are both zero, and variances
$\left<\hat{a}^{2}\right>=\hat{q}$,$\left<\hat{a}'^{2}\right>=\hat{r}$)
according to their definition, this is because,
$\hat{q}-\widehat{p'}$ or $\hat{r}-\widehat{p'}$ may get negative.
\section{Derivation of $\frac{{\rm d}\mathcal
{V}(p)}{{\rm d}p}|_{p\rightarrow 1}$} \label{app:Deriv}
The Franz-Parisi potential $\mathcal {V}(p)$ is obtained through a
Legendre transform of $f(x)$, i.e., $\mathcal {V}(p)=f(x)-xp$. The
overlap $p\equiv\mathbf{J}\cdot\mathbf{w}/N$ is related to the
coupling field by $\frac{{\rm d}f(x)}{{\rm d}x}=p$. Since the
potential curve may lose its concavity, one has to solve numerically
the saddle-point equations by fixing $p$ and searching for
compatible coupling field $x$ (by using the secant method). If a
solution of $x$ is found for a given $p$, then we have
$x=-\frac{{\rm d}\mathcal {V}}{{\rm d}p}$ at this value of $p$.
Because $d=(1-p)/2$, $x$ is also equal to $\frac{1}{2}\frac{{\rm
d}\mathcal {V}}{{\rm d}d}$.

The derivative of the Franz-Parisi
potential with respect to the overlap $p$ is given by:
\begin{equation}\label{slope}
\begin{split}
    \frac{{\rm d}\mathcal
{V}(p)}{{\rm d}p}&=-\hat{p}+\alpha\frac{\partial}{\partial p}\int
D\omega\int Dt H^{-1}(\tilde{t})\int_{\tilde{t}}^{\infty}Dy\ln
H(h(\omega,t,y))\\
&=-\hat{p}+\frac{\alpha}{\sqrt{(1-q)(1-r)}}\int D\omega\int Dt
\mathcal
    {R}(\tilde{t})\mathcal
    {R}(h(\omega,t,y=\tilde{t})).
\end{split}
\end{equation}
Note that when $p\rightarrow1$, $r$ will get close to $p$ but
smaller than $p$, and $p'\simeq q$, which is observed in numerical
simulations and can be understood from the definition of these order
parameters. Therefore, in the limit $p=1-\epsilon\rightarrow1$, the
second term in the right-hand side of Eq.~(\ref{slope}) is $\alpha
C_{p}\epsilon^{-1/2}$ with $C_p=\frac{1}{\sqrt{\pi(1-q)}}\int
Dt\mathcal {R}(\tilde{t})$. The expression of $\hat{p}$ as a
function of $\epsilon$ can be deduced from Eq.~(\ref{SDEc}). Using
the fact that $p\rightarrow 1$ implies that
$\hat{p}\rightarrow\infty$, and the identity $\tanh(x)=1-2e^{-2x}$
($x\gg0$), one finally gets
$\hat{p}=\hat{r}-\frac{1}{2}\ln\frac{\epsilon}{2}+\frac{1}{2}\ln\int
Dz_1\int
Dz_3\frac{e^{\sqrt{\hat{q}-\widehat{p'}}z_1-\sqrt{\widehat{p'}}z_3}}{\cosh\left(\sqrt{\hat{q}-\widehat{p'}}z_1+\sqrt{\widehat{p'}}z_3\right)}$.
In the above derivations, we have used the fact that
$\frac{1-p}{1-r}=1/2$ in the limit $p\rightarrow1$ based on
Eqs.~(\ref{SDEc}) and~(\ref{SDEe}). Taken together, one arrives at
the slope of $\mathcal {V}(p)$ at $p=1$:
\begin{equation}\label{slope2}
 \frac{{\rm d}\mathcal
{V}(p)}{{\rm
d}p}|_{p\rightarrow1}=\frac{1}{2}\ln\frac{\epsilon}{2}+C+\alpha
C_{p}\epsilon^{-1/2}.
\end{equation}
\section{Proof of $\mathcal {V}(p\rightarrow 1)=0$}
\label{app:Proof}
At $p=1$, the Franz-Parisi potential can be expressed as:
\begin{equation}\label{VFP02}
\begin{split}
    \mathcal {V}(p) &=-p\hat{p}+p'\widehat{p'}+\alpha\int D\omega\int DtH^{-1}(\tilde{t})\int_{\tilde{t}}^{\infty}Dy\ln H(h(\omega,t,y))\\
    &+\int
    D\mathbf{z}(2\cosh\hat{a})^{-1}\Biggl[e^{\hat{a}}\ln2\cosh(\hat{a}'+\hat{p}-\widehat{p'})+e^{-\hat{a}}\ln2\cosh(\hat{a}'-\hat{p}+\widehat{p'})\Biggr].
    \end{split}
\end{equation}
Note that
$h(\omega,t,y)=-\frac{1}{\sqrt{1-r}}\left(\sqrt{1-q}y+\sqrt{q}t\right)\rightarrow-\infty$
when $y>-\sqrt{\frac{q}{1-q}}t$. Hence the $\alpha$-dependent term
disappears. The last term becomes
\begin{equation}\label{VFP02}
\begin{split}
    \int
    &D\mathbf{z}(2\cosh\hat{a})^{-1}\Biggl[e^{\hat{a}}\ln2\cosh(\hat{a}'+\hat{p}-\widehat{p'})+e^{-\hat{a}}\ln2\cosh(\hat{a}'-\hat{p}+\widehat{p'})\Biggr]\\
    &=\int Dz_1\int Dz_3
    (2\cosh\hat{a})^{-1}\left[e^{\hat{a}}(\sqrt{\widehat{p'}}z_3+\hat{p}-\widehat{p'})+e^{-\hat{a}}(-\sqrt{\widehat{p'}}z_3+\hat{p}-\widehat{p'})\right]\\
&=\hat{p}-\widehat{p'}+\widehat{p'}\left[1-\int Dz_1\int Dz_3\tanh^{2}\Bigl(\sqrt{\hat{q}-\widehat{p'}}z_1+\sqrt{\widehat{p'}}z_3\Bigr)\right]\\
    &=\hat{p}-q\widehat{p'}.
    \end{split}
\end{equation}
Collecting the above results, one arrives at $\mathcal
{V}(p\rightarrow1)=0$.


\end{document}